\documentclass{article}
\usepackage{graphicx}
\usepackage{subcaption}
\usepackage{amsthm}
\usepackage{amsmath}
\usepackage{amsfonts}
\usepackage[affil-it]{authblk}

\usepackage{pict2e}
\usepackage{graphicx}
\usepackage{multirow}
\usepackage{svg}

\usepackage{tikz}
\usepackage{tikz-3dplot}
\usepackage{quantikz}
\usepackage{subcaption}

\usepackage{url}

\newtheorem{definition}{Definition}

\newtheorem{theorem}{Theorem}

\title{Solving Segment Display Problems Using Quantum Grover's Search Algorithm}

\author[1,2]{Shanyan Chen}
\author[2]{Ali Al-Bayaty}
\author[2]{Xiaoyu Song}
\author[2]{Marek Perkowski}

\affil[1]{College of Information Engineering, Capital Normal University, China}
\affil[2]{Department of Electrical and Computer Engineering, Portland State University, USA}
\date{}

\begin{document}

\maketitle

\begin{abstract}
    This paper introduces a new Boolean-based methodology for constructing Segment Display Problems (SDPs) in the quantum domain and solving them using Grover's quantum search algorithm.
    In the classical domain, the SDPs are typically solved using various techniques, such as human deduction, heuristic search, and methods for solving Boolean satisfiability (SAT) and constraint satisfaction problems (CSPs) that are based on different problem design models. In this paper, our newly introduced methodology proposes a quantum-based approach for solving such SDPs, by 
    building their quantum oracle using binary reversible circuits and our previously proposed step-decreasing structures shaped operators (Stesso). To demonstrate the usability of this proposed method, we experimentally solve an SDP instance of the matchstick problem using Grover's algorithm with a noisy simulated quantum computer implemented in Qiskit.
\end{abstract}


\footnotetext[1]{Email of corresponding author: sychen@cnu.edu.cn}
\footnotetext[2]{Email of authors: shanyan@pdx.edu, albayaty@pdx.edu, songx@pdx.edu, h8mp@pdx.edu}

\section{Introduction}

As a niche but intriguing area of logical and computational problems, Segment Display Problems/Puzzles (SDPs)~\cite{wiki:SDP,arcturus2011pub} leverage the universal, constrained visual grammar of the seven-segment display~\cite{genevra2013effective} (or similar configurations) to pose challenges that range from simple visual trickery to complex Constraint Satisfaction Problems (CSPs)~\cite{bulatov2011complexity}. Certain types of SDPs are known to be NP-hard or PSPACE-complete~\cite{uehara2023computational}.

In general, SDPs often serve as pedagogical tools for teaching digital logic, Boolean algebra, and fundamental algorithm design~\cite{parhami2009puzzling,falkner2010puzzle}. Numerous SDPs have even appeared in the World Puzzle Championship (WPC) and related unofficial competitions~\cite{wiki:WPC}.
Solving these SDPs requires deducing the correct state (lit or unlit) of each segment under a set of given constraints, partial information, or required arithmetic operations. This domain offers a tangible framework for comparing the approaches of classical computing, which relies on systematic search~\cite{chen2025enigmata} and heuristic optimization~\cite{desale2015heuristic}, with quantum computing. This explores the potential for polynomial speedups \cite{ronnow2014defining} in specific problem types by leveraging quantum-mechanical principles, such as superposition and entanglement.

For this reason, the goal of our paper is to introduce a state-of-the-art methodological approach in the quantum domain for: (i) implementing a classical SDP using quantum Boolean oracles and circuits, (ii) representing these quantum oracle and circuits using seven-segment codes (SSC), binary decoders, binary comparators, and cryptarithmetic circuits, (iii) optimizing the quantum circuits of SC and binary decoders using our recently proposed ``step-decreasing structures shaped operators (Stesso)" approach in \cite{chen2025stesso}, and (iv) solving the
quantum-based SDPs using Grover's algorithm \cite{grover1996fast,mandviwalla2018implementing} in the context of matchstick problems~\cite{danek2016solving}.

In quantum computing, there are several well-known quantum search algorithms with significant efficiency improvements for finding correct solutions in polynomial speedup compared to the classical search algorithms.
For instance, these quantum search algorithms are Grover's algorithm, Quantum Random Walk algorithm \cite{xia2019random}, and Boolean-Hamiltonians Transform for Quantum Approximate Optimization Algorithm (BHT-QAOA) \cite{al2024bht}.
In our research, we utilized Grover's algorithm to find all solutions for SDPs, due to the following three facts: (i) Grover's algorithm quadratically speeds up the searching mechanism in the evaluation complexity of $O(\sqrt {N})$, where $N = 2^n$ and $n$ is the number of input qubits, (ii) Grover's algorithm is able to search for solutions for both quantum Boolean and phase oracles; hence, an SDP can be easily implemented as a Boolean oracle using reversible gates, and (iii) our recently introduced diffuser \cite{al2024concept} is able to find all correct solutions for a Boolean oracle regardless of its different logical structures, as compared to the standard diffuser of Grover's algorithm that is limited to a defined set of logical structures \cite{mandviwalla2018implementing,grover1998framework}. Therefore, Grover's algorithm with a Boolean oracle and our diffuser can efficiently solve such segment display problems (SDPs), which are considered as a combination of constraint satisfaction problems (CSPs) \cite{brailsford1999constraint} and Boolean satisfiability (SAT) problems \cite{massacci2000logical,utomo2017solving}.

In this paper, a general method for modeling and solving SDPs is introduced, and an example matchstick problem is simulated experimentally using the IBM Qiskit quantum library \cite{fadillahintroduction}. The simulated results obtained with Grover's algorithm are correct, demonstrating that our methodological approach can be used to implement and solve various SDPs successfully and can even be extended to other CSPs.



\section{Background and Related Work}

This section presents the background and related work on segment display problems and on classical and quantum approaches to solving problems.

\subsection{Segment display problems}

Segment Display Problems/Puzzles (SDPs)~\cite{wiki:SDP} are a popular category of recreational mathematics and logic problems that utilize the familiar visual representation of digits found in digital clocks and calculators, which are tied to the display technology itself. Commonly used segment displays can be classified into seven-segment, fourteen-segment, and sixteen-segment types by the number of segments~\cite{displays2025chap}. 



\begin{figure}[htbp]
    \centering
    \includegraphics[width=0.85\linewidth]{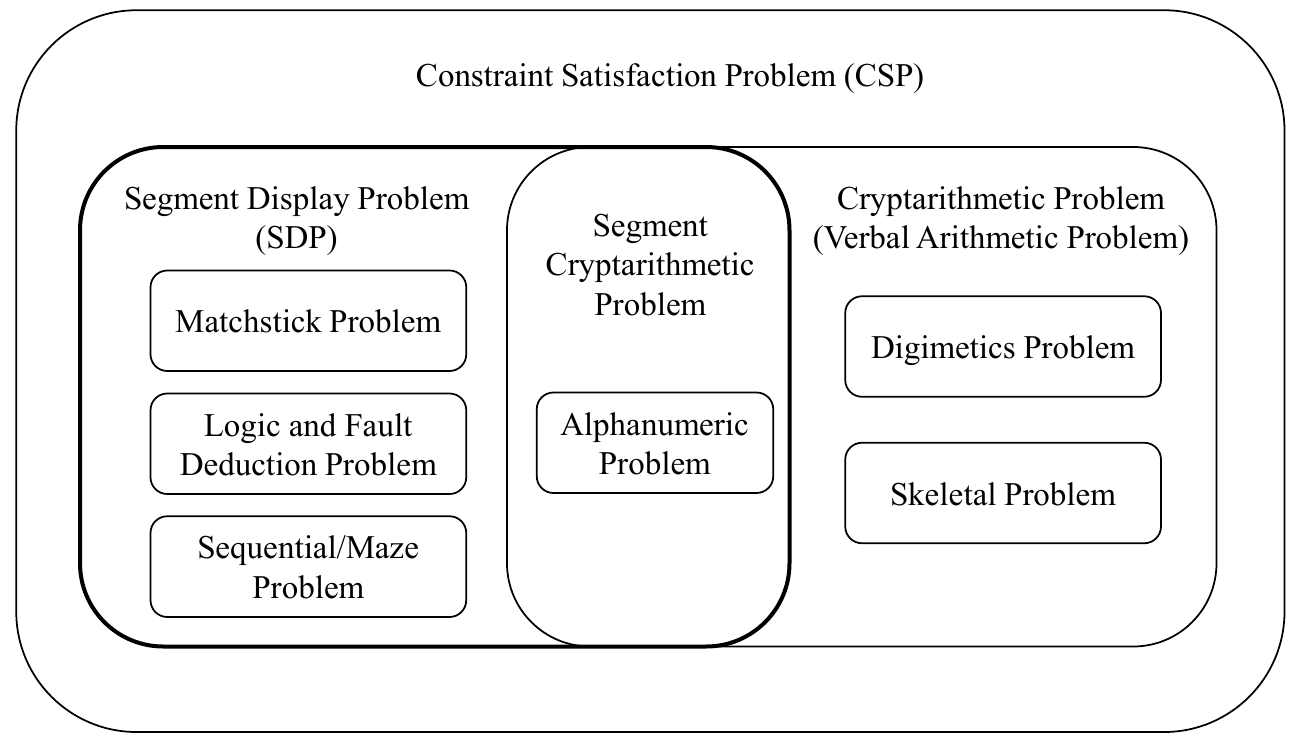}
    \caption{Group relationships among constraint satisfaction problems (CSPs), segment display problems (SDPs), and cryptarithmetic problems, along with some examples.}
    \label{fig:relation}
\end{figure}

Therefore, these SDPs challenge users to manipulate the segments, often with constraints, to form valid numbers, words, or solve arithmetic problems. So, SDPs are also constraint satisfaction problems (CSPs)~\cite{brailsford1999constraint}. According to the relevant geometric and arithmetical constraints, the relationships among CSPs, SDPs, and cryptarithmetic problems~\cite{minhaz2014solution,yang2020solving} can be classified as shown in Figure~\ref{fig:relation}.

Due to the number of segments and the supported configurations (numbers and some letters), the seven-segment display~\cite{genevra2013effective} is the most commonly used among segment displays.
So, the commonly mentioned SDPs in practical tools and benchmarks refer to problems based on seven-segment displays, such as matchstick problems~\cite{danek2016solving}, logic and fault deduction problems~\cite{wiki:WPC}, and sequential/maze problems~\cite{SegMaze,kurokawa2014picture} in Figure~\ref{fig:relation}. These are not standard cryptarithmetic problems, also named as verbal arithmetic problems~\cite{jerman1972predicting,mateus2021development}, that involve only letters, such as digimetics problems~\cite{MK2025Digi}, and skeletal problems~\cite{dudeney2016536}. 

Unlike other examples of SDPs based on seven-segment displays, there are different types of problems typically based on fourteen-segment or sixteen-segment displays, termed as the segment cryptarithmetic problem~\cite{nareyek2001send}. This segment cryptarithmetic problem is a hybrid alphanumeric problem involving letters, which belongs to the overlap area between the SDP and the cryptarithmetic problem, as shown in Figure~\ref{fig:relation}. 


\subsection{Classical approaches for solving SDPs}

As a classical logical problem, the SDP is a fascinating combinatorial and computational challenge that has been tackled using a variety of classical approaches.
These approaches can range from human deduction, heuristics, and methods for solving Boolean satisfiability (SAT)~\cite{utomo2017solving,alyahya2022structure} and CSPs \cite{brailsford1999constraint,bulatov2011complexity}.

For most human-designed logic problems, such as ``Two Clocks" or ``Digital Scale" in WPC~\cite{wiki:WPC}, solving relies on intuitive strategies for human deduction and on search algorithms since the rules of these problems constrain their solution space. 
So the heuristic search method \cite{desale2015heuristic}, which aims to find a solution by exploring the space of possible states using domain-specific strategies, can also be applied to human-designed SDPs.
For instance, the greedy search algorithm \cite{wilt2010comparison} finds the next variable based on intermediate constraint satisfaction in specific equations, and the A* algorithm \cite{hart1968formal} utilizes an admissible heuristic technique to guide the search. If the SDPs are physically present, algorithms in image processing and computer vision that use image features, edge detection, and shape analysis to digitize the problem elements before applying logical solvers are efficient \cite{fadhil2025automatic}.

For automated solvers, SDPs are often modeled as Boolean satisfiability (SAT) problems or CSPs. If an SDP is modeled as a Boolean SAT problem, it means the SDP is translated into a massive Boolean formula, where variables indicate whether a segment is lit.
Then modern SAT solvers~\cite{martins2012overview,alouneh2019comprehensive} are highly effective tools for determining whether there exists an assignment of true/false to these variables that makes the entire formula true.
If an SDP is modeled as a CSP, it means the SDP is translated into finding a valid configuration of segment displays that satisfies constraints, e.g., valid mathematical equations, word formation, etc. Hence, all approaches for CSPs can also be applied to SDPs, such approaches are: (i) the arc-consistency algorithm \cite{van1992generic} to ensure the satisfaction of binary constraints, (ii) Knuth's Algorithm X \cite{knuth2000dancing} affected by the limitations on involved covering sets, and (iii) the backtracking algorithm \cite{van2006backtracking} in case a violation of constraints is detected.

All the above-mentioned classical approaches are still designed and implemented on classical computer systems; however, their efficiency is limited by hardware computing power and software design. For this reason, compared with these classical approaches, our research in this paper introduces a new general methodology to solve SDPs using Grover's algorithm in the quantum domain.





%

\subsection{Quantum approaches for specific problem solving}

Compared with classical approaches to problem solving, quantum approaches are still largely theoretical or in early experimental stages, often focusing on mapping specific problems onto quantum computing paradigms.

For large-scale problems, current practical approaches often combine classical pre- or post-processing with quantum computation for the most challenging parts of the problem, such as the combinatorial problems~\cite{wurtz2024solving}. This hybrid quantum-classical method~\cite{bar2023quantum,qazi2025quantum} breaks down large problems into modules that can run on contemporary Noisy Intermediate-Scale Quantum (NISQ) devices~\cite{callison2022hybrid}. 

For some complex CSPs without large scale, they can be reformulated as Quadratic Unconstrained Binary Optimization (QUBO) problems~\cite{punnen2022quadratic} and solved using quantum annealers \cite{jiang2022solving,jiang2023classifying}. This method aims to find the global minimum ground state (the correct solution) by leveraging quantum tunneling to explore a solution space efficiently.

For certain problem structures that can be converted into quantum search problems, all quantum search algorithms can be applied to solve these problems, such as Grover's algorithm \cite{mandviwalla2018implementing}, Quantum Random Walk algorithm \cite{xia2019random}, and Boolean-Hamiltonians Transform for Quantum Approximate Optimization Algorithm (BHT-QAOA) \cite{al2024bht}. 

Take a maze problem~\cite{SegMaze,kurokawa2014picture} as an example, several researchers have used quantum search algorithms to solve it. Kumar et al.~\cite{kumar2013quantum} use Grover's algorithm to solve the quantum search problem converted from a maze problem. Different from Grover's algorithm, Caruso et al.~\cite{caruso2016fast} demonstrate how a quantum walker can efficiently reach the output of a maze by partially suppressing the presence of interference. Still using quantum walk, Matsuoka et al.~\cite{matsuoka2025mathematical} recently introduced a mathematical framework for maze solving. Although all these works proved that quantum search algorithms are more efficient than classical ones for solving 
maze problems, they are restricted to specific maze problems. 

Notice that a sequential/maze problem can be presented by segment logic, i.e., a sequential/maze problem is an SDP as demonstrated in Figure~\ref{fig:relation}.
Thus, unlike previous work, our paper introduces a new methodological approach for solving SDPs in the quantum domain, which is more general and easier to extend to other CSPs. To demonstrate the usability of our proposed method, we apply it to solve the matchstick problem (as an SDP) and simulate it on Qiskit.
The outputs of our method always have promising results, meaning that this new methodological approach can be further applied to solve various SDPs and other types of CSPs.

\section{Methods for a High-Level Oracle Design}

In this paper, our methodology for the oracle design of the segment display problem (SDP) is categorized into a high-level and a logic level. The high-level means that an oracle is designed using several blocks, such as the segment code (SC) verifier, segment code to binary code decoder (SC-BCD), and arithmetic units. In comparison, the logic level introduces the overall design from the above-mentioned high-level blocks, as layout-aware blocks for a real quantum computer. In this section, we introduce our methodology for a high-level oracle design.

\begin{figure}[htbp]
    \centering
    \includegraphics[width=0.8\linewidth]{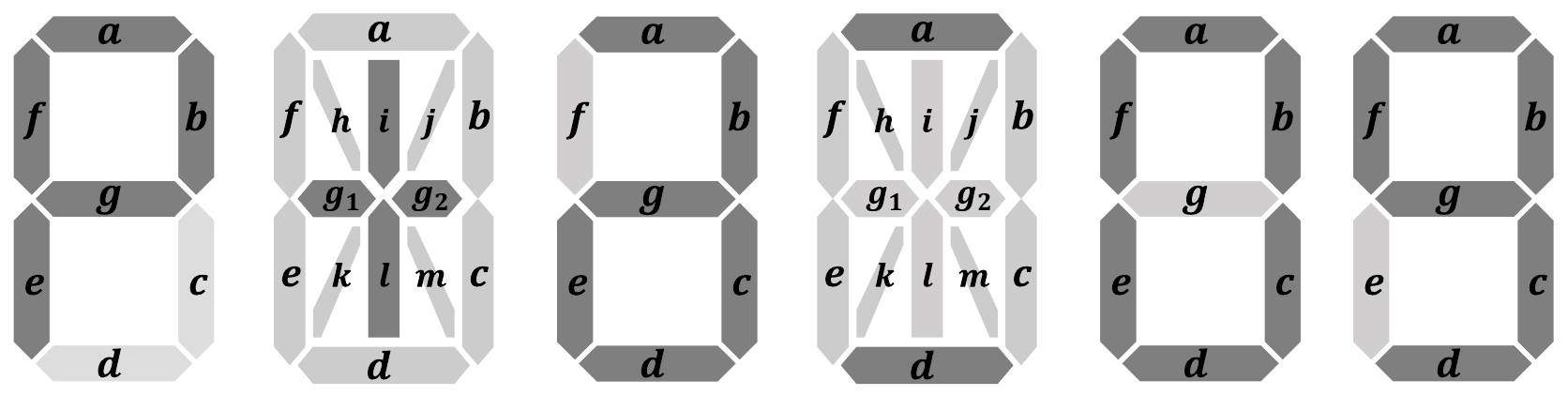}
    \caption{An example of an SDP for adding two one-digit numbers. A seven-segment display presents a one-digit number using a 7-bit binary vector $[a,b,c,d,e,f,g]$, while a fourteen-segment display presents an operation using a fourteen-bit binary vector $[a,b,c,d,e,f,g_1,g_2,h,i,j,k,l,m]$.}
    \label{fig:SDPexp}
\end{figure}

Since an SDP is typically based on the $n$-segment displays, where $n$ denotes the segment number, we can analyze the suitable construct model of an SDP beginning with a simple example as shown in Figure~\ref{fig:SDPexp}. In Figure~\ref{fig:SDPexp}, the representation of the mathematical equation consists of two one-digit numbers and one two-digit number (presented by four physical seven-segment displays) and the addition and equal operations (presented by two fourteen-segment displays). Note that the operations in an SDP can be represented by abstract binary codes, yielding a simpler model, especially when the number of all used operations is less than three or some operations are fixed. 

Obviously, the mathematical model of an SDP includes a variable set $Q$ with domain $D$ that satisfies the constraints $C$ derived from its operations and from valid configurations. Notice that the design of a general problem for real computers, whether in the classical or quantum domain, not only contains the design of its mathematical model theoretically, but also includes the design of its realistic model adapted from the final applied domain (specifically the quantum domain in this paper).





Thus, regarding the specific example of an SDP, our methodology is consistent with the design ideas of general classical problems, which are generally divided into three parts: (i) encoding an SDP, (ii) defining the correctness of the solution to the SDP, and (iii) designing the Boolean oracle that is constructed from classical Boolean gates. The first two parts (i) and (ii) correspond to the mathematical model of an SDP. The final part (iii) corresponds to the realistic model of an SDP in the quantum domain.

\subsection{Encoding an SDP}

For an $n$-segment display, the total number of supported configurations, such as numbers and letters, increases when the segment number $n$ increases. Considering the number of segments and the supported configurations, common $n$-segment displays include seven-segment displays, fourteen-segment displays, and sixteen-segment displays, where $n$ equals seven, fourteen, and sixteen, respectively. Notice that the seven-segment display is the most common $n$-segment display, which is used to represent a digit number in Figure~\ref{fig:SDPexp}. 

A $n$-segment display consists of $n$ independently controlled LED-based segments, and the state (bright/dark) of each LED segment is represented by the corresponding independent binary variable for each segment, denoted by an $n$-bit binary code vector $[q_1,\cdots,q_n]$. Usually there is specific notation for an standard $n$-segment display, such as a 7-bit binary code vector $[a,b,c,d,e,f,g]$ for an seven-segment display and a fourteen-bit binary code vector $[a,b,c,d,e,f,g_1,g_2$, $h,i,j,k,l,m]$ for an fourteen-segment display in Figure~\ref{fig:SDPexp}. For ease of representation, we use $abcdefg$ to denote the 7-bit binary code vector $[a,b,c,d,e,f,g]$. All other vectors in this paper follow the same way of representation.




Thus, the variable $q_{s}$ for an $n$-segment display is defined in Definition~\ref{def:vset_seg}, with its domain $D_{seg}$ in Definition~\ref{def:domain_seg}.

\begin{definition}
    \label{def:vset_seg}
    The segment display variable $q_{s}$ for an $n$-segment display is an $n$-binary vector for all physical segments based on its standard labeled variables of each segment, denoted as $q_{s} = q_{s_1} \cdots q_{s_n}$.
\end{definition}

\begin{definition}
    \label{def:domain_seg}
    The domain $D_{seg}$ for an $n$-segment display is the Cartesian product $\bigtimes$ of all nonempty domains $D_{si}$ with respect to all physical segment variables $q_{s_i}$ of the vector $q_s$, i.e., $D_{seg} = D_{si}^{n}$, $D_{si} = \{ \ket{0}, \ket{1} \}$ where $\ket{0}$ and $\ket{1}$ represent dark and bright of an LED segment respectively.
\end{definition}

For an $n$-segment display, some of its configurations can present a valid number or a letter. Therefore, another variable $q_{d}$ is required to stand for the valid digit number obtained from the visual presentation of an $n$-segment display, as stated in Definition~\ref{def:vset_dig}. Note that a letter in an SDP, especially in a segment cryptarithmetic problem, typically yields a unique number of solutions. For this reason, the elements of the domain $D_{dig}$ in Definition~\ref{def:domain_dig} of the variable $q_{d}$ are integer numbers.

\begin{definition}
    \label{def:vset_dig}
    The digit position variable $q_{d}$ that indicates a one-digit number is denoted as an $i$-bit binary code vector, where $2^{i}-1$ is no less than the number of all elements of its corresponding domain $D_{dig}$.
\end{definition}

\begin{definition}
    \label{def:domain_dig}
    The domain $D_{dig}$ is the set of natural numbers from 0 to 9, i.e., $D_{dig} = [0, 9]$.
\end{definition}

As shown in Figure~\ref{fig:SDPexp}, an SDP is usually composed of numbers (from 0 to 9) and arithmetic operations (such as addition, subtraction, and equality, etc). Therefore, the encoding of an SDP can be divided into two parts, one representing the corresponding numbers and the other representing the operators.

So far, the previous definitions can encode the numbers of an SDP. In this paper, an $n$-bit binary code vector $q_{s}$ is termed the ``Segment Code (SC)''. Its corresponding states are termed ``SC-State", which corresponds to the domain $D_{seg}$. 

Then, for the encoding of the operators of an SDP, we use a variable $q_o$ to present each operator involved in an SDP. There are two ways to represent an operation in an SDP, depending on the number of total operators used and the presence of fixed operators, i.e., physical representation based on $n$-segment displays and abstract binary code vector representation. Here, we use an abstract definition of an operator in an SDP that is more general and applicable across different code formats, as stated in Definition~\ref{def:vset_op} along with its corresponding domain in Definition~\ref{def:domain_op}.

\begin{definition}
    \label{def:vset_op}
    The operation variable $q_{o}$ that stands for an operation is denoted as a $k$-bit binary code vector, where $2^{k}-1$ is no less than the number of actual used elements of its corresponding domain $D_{op}$ in an SDP.
\end{definition}

\begin{definition}
    \label{def:domain_op}
    The domain $D_{op}$ is the set of all supported operators for operation variable $q_o$, i.e., $D_{op} = \{ +, -, \times, \div,  <, =, > \}$.
\end{definition}

Notice that the lengths of a digit position variable $q_{d}$ and an operation variable $q_o$ are flexible based on the possible valid number of elements in $D_{dig}$ and $D_{op}$, respectively.
For example, if $D_{dig_j} = [0,9]$ without constraints related to the $j$-th digit position in a specific SDP, then the $j$-th digit position variable $q_{dj}$ can be a 4-bit binary code vector, i.e., $q_{dj} = q_{dj_1}q_{dj_2}q_{df_3}q_{dj_4}$.

Now, based on the definition of the different variables $q$ for segment displays, digit numbers, and operators, the whole variable set $Q$ of an SDP can be defined as the set of all these variables in Definition~\ref{def:vset}. The domain set $D$ of the variable set $X$ is defined analogously in Definition~\ref{def:domain}.

\begin{definition}
    \label{def:vset}
    The variable set $Q$ of an SDP, based on the digits presented by $m$ $n$-segment displays and $l$ operations presented by abstract binary code, each digit position variable $q_d$ that represents a number or letter in the display, and the operation variables $q_o$ for all involved operations, i.e., $Q = Q_{seg} \cup Q_{dig} \cup Q_{op}$, $Q_{seg} = \{ q_{sj} | 1 \leq j \leq m, q_{sj}  = q_{sj_1} \cdots q_{sj_n} \}$, $Q_{dig} = \{ q_{dj} | 1 \leq j \leq m \}$, $Q_{op} = \{ q_{ok} | 1 \leq k \leq l \}$, where $j$ is index of an $n$-segment display and $k$ is the index of involved operations in an SDP.
\end{definition}

\begin{definition}
    \label{def:domain}
    The domain set $D$ of an SDP consists of all nonempty domains with respect to all physical segment variables $q_s$ in set $Q_{seg}$, digit position variables $q_d$ in set $Q_{dig}$, and operation variables $q_o$ in set $Q_{op}$, i.e., $D = \bigcup_{j=1}^{m}{(D_{seg_{j}} \cup D_{dig_{j}})} \cup \bigcup_{k=1}^{l}D_{op_{k}}$ where $j$ is index of an $n$-segment display and $k$ is the index of involved operations in an SDP. 
\end{definition}


In the following sections, we use $i$, $j$, and $k$ to indicate the index of a segment in an $n$-segment display, the index of a segment display for all $m$ segment displays in an SDP, and the index of an operation used (involved) in an SDP, respectively. 

\subsection{Defining the correctness of the solution to the SDP}

For an SDP, the problem challenges users to manipulate the segments, often with constraints, to form valid numbers, words, or solve arithmetic problems. Take the example shown in Figure~\ref{fig:SDPexp}, the correctness of the solution to a problem must meet four necessary conditions as follows.
\begin{itemize}
    \item Condition 1: the correctness of  SC-State.
    \item Condition 2: the correctness of the cryptarithmetic operator.
    \item Condition 3: the correctness of the cryptarithmetic equation.
    \item Condition 4: the number of changed segments, which we termed $K$.
\end{itemize}

The first two conditions originate from the assumption that both the numbers and the operators are correct in an SDP.
If the numbers and operators in the problem are incorrect, the equation in the problem is meaningless.
The third condition is then derived.
Notably, all segment states in the problem may be changed.
Therefore, this fact derives the fourth condition.
If the above four conditions hold, then there exists an SDP without additional constraints solution for $K$ segments.
If there are additional constraints of an SDP that exist, these additional constraints are also required to be held, all classified as the fifth condition.
\begin{itemize}
    \item Condition 5: additional constraints: such as the independent SC-States for different letters in the alphanumeric problem, sequential relation of different displays, etc  
\end{itemize}

According to the conditions mentioned above, all constraints of an SDP can be categorized into three types: geometric constraints inside one $n$-segment display, among different segment displays, and cryptarithmetic constraints.

For the geometric constraints inside an $n$-segment display, termed inner geometric constraints $C_{gin}$, these constraints are the relations between the physical segment variable $q_s$ and its corresponding digit position variable $q_d$, as illustrated in Definition~\ref{def:cons_inner}.

\begin{definition}
    \label{def:cons_inner}
    Inner geometric constraints of an SDP are the constraints inside one $n$-segment display, denoted as $C_{gin}:D_{seg} \bigtimes \{ \ket{0}, \ket{1}\} \bigtimes D_{dig} \rightarrow \{ 0, 1\}$.
\end{definition}

Obviously, the correctness of SC-State (Condition 1) is an inner geometric constraint $C_{gin}$ in Definition~\ref{def:cons_inner}. For an $n$-segment display, there are only a few correct SC-States for all possible SC-States, e.g., a maximum of ten correct SC-States $s_{num}$ for valid digit numbers and twenty-six correct SC-States $s_{letter}$ for valid letters without other constraints. Thus, the correctness of SC-States can be described as a sparse function $f_{gin}$ that has only a few values that are one. Another inner geometric constraint $C_{gin}$ is the meaning of an SC-State, which can usually be described as a bijection function $f_{gin}$ that maps a segment variable $q_s$ to a digit position variable $q_d$. Notice that the meaning of an SC-States can be a part of independent SC-States listed in Condition 5. Usually, the inner geometric constraints $C_{gin}$ consist of only the previous two examples. Therefore, the Theorem~\ref{th:gin_f} derived.

\begin{theorem}
    \label{th:gin_f}
    The inner geometric constraints $C_{gin}:D_{seg} \bigtimes \{ \ket{0}, \ket{1}\} \bigtimes D_{dig} \rightarrow \{ 0, 1\}$ always can be described by the function $f_{gin}$ that is a sparse function related to $D_{seg}$ and $\{ \ket{0}, \ket{1}$ or a bijection function related to $D_{seg}$ and $D_{dig}$, such as the following equation:
    \begin{equation*}
    \begin{aligned}
        & \forall q_s. ~ q_s = s_{num} \vee s_{letter} \Leftrightarrow f_{gin}(q_s) = \ket{1}\\
        &
        \forall q_{s1}~q_{s2}~q_{d1}~q_{d2}. ~ 
        q_{s1} \neq q_{s2} \wedge
        f_{gin}(q_{s1}) = q_{d1} \wedge f_{gin}(q_{s2}) = q_{d2} \Rightarrow 
        q_{d1} \neq q_{d2}\\
        &
        \forall q_d. ~\exists q_s.~
        f_{gin}(q_s) = q_d
    \end{aligned}
    \end{equation*}
\end{theorem}

For the geometric constraints among different segment displays, termed amid geometric constraints $C_{gamid}$, these constraints are the relations among different segment variables $q_{sj}$ in a subset of $Q_{seg}$, as defined in Definition~\ref{def:cons_amid}.

\begin{definition}
    \label{def:cons_amid}
    Amid (inter) geometric constraints of an SDP are the constraints between several $n$-segment displays, denoted as $C_{gamid}:(\bigtimes_{j \in s}^{s \subset [1,m]}{D_{seg_j}})$ $ \bigtimes \mathbb{N} \rightarrow \{0,1\}$.
    
\end{definition}

One example of the $C_{gamid}$ is related to the condition about the changed $K$ segments (Condition 4). Considering the number $K$ of changed segments, this condition is always related to the Hamming distance (HD) between the SC-States of two $n$-segment displays, which counts the number of positions where their corresponding states differ. For this reason, the domain $\mathbb{N}$ of number $K$ of changed segments is involved in the definition of $C_{gamid}$ (Definition~\ref{def:cons_amid}). Another example of $C_{gamid}$ is to count all ones in the SC-States (the lighted LED segments) of all segment displays, usually more than two segment displays, on the left or right side of an operation in an SDP. Moreover, the sequential relation among the different displays mentioned in Condition 5 is clearly an amid geometric constraint, also related to HD. Therefore, Theorem~\ref{th:gamid_f} is obtained.
\begin{theorem}
    \label{th:gamid_f}
    The amid geometric constraints $C_{gamid}$ can always be described by the Hamming weight function based on the HD, along with a comparison.
\end{theorem}

For the cryptarithmetic constraints, these constraints are the relations among the arithmetical equations consisting of the operation variables $q_{o}$ and the digit position variables $q_{d}$, as described in Definition~\ref{def:cons_arith}.

\begin{definition}
    \label{def:cons_arith}
    Cryptarithmetic constraints of an SDP are the constraints of operators and formed cryptarithmetic equations, denoted as $C_{art}:(\bigtimes_{j \in s}^{s \subset [1,m]}{D_{dig_j}})$ $(\bigtimes_{k \in s}^{s \subset [1,l]}{D_{op_k}}) \rightarrow \{ 0,1\}$.
\end{definition}

If the operator in an SDP is presented by a segment display, such as the example in Figure~\ref{fig:SDPexp}, the second condition about the correctness is an inner geometric constraint $C_{gin}$. Since the definition of the operation variable $q_o$ in Definition~\ref{def:vset_op} is applicable across different code formats, the correctness of the operators (Condition 2) becomes a cryptarithmetic constraint $C_{art}$ and a special case of a general inner geometric constraint $C_{gin}$ at the same time. Notice that the special case of a general inner geometric constraint $C_{gin}$ mentioned here is not the exact Definition~\ref{def:cons_inner} because of the different involved domain $D_{op}$. The reason why the second condition can be categorized as the special case of a general inner geometric constraint $C_{gin}$ is that it remains a ``geometric configuration" (actual requirements of format) of the operation variable $q_o$ in a sense.
Obviously, the correctness of the cryptarithmetic equation (Condition 3) is also a cryptarithmetic constraint $C_{art}$.

Based on the three types of constraints $C_{gin}$, $C_{gamid}$, and $C_{art}$ (Definition~\ref{def:cons_inner}, \ref{def:cons_amid}, and \ref{def:cons_arith}), the definition of constraint set $C$ is illustrated in Definition~\ref{def:cons_all}.
\begin{definition}
    \label{def:cons_all}
    The constraint set $C$ that each constraint involves a subset of variables and restricts their admissible combinations of values, is the set of inner geometric constraints $C_{gin}$, amid geometric constraints $C_{gamid}$, and cryptarithmetic constraints $C_{art}$, denoted as $C = \{ C_{gin}, C_{gamid}, C_{art} \}$. 
\end{definition}

\subsection{Designing a Boolean oracle}

Based on definition of variable set $Q$ (Definition~\ref{def:vset}) with its domain set $D$ (Definition~\ref{def:domain}) and the constraint set $C$ (Definition~\ref{def:cons_all}), an SDP can be defined as a tripe $\langle Q,D,C \rangle$ in Definition~\ref{def:SDP}.

\begin{definition}
    \label{def:SDP}
    A segment display problem (SDP) is a type of constraint satisfaction problem (CSP) that uses the specific structure of segment displays (typically the seven-segment display) as its domain of constraints, denoted as a triple $\langle Q,D,C\rangle$, where $Q$ is the variable set, $D$ is the domain set with respect to $Q$, and $C$ is the constraint set.
\end{definition}

In quantum computing, the design and implementation of hardware quantum layouts and circuits essentially affect the efficiency of quantum search algorithms. 
For instance, in Grover's algorithm, the essential part of designing and implementing a quantum circuit mainly depends on the design of the quantum oracle.

\begin{figure}[htbp]
    \centering
    \includegraphics[width=0.85\linewidth]{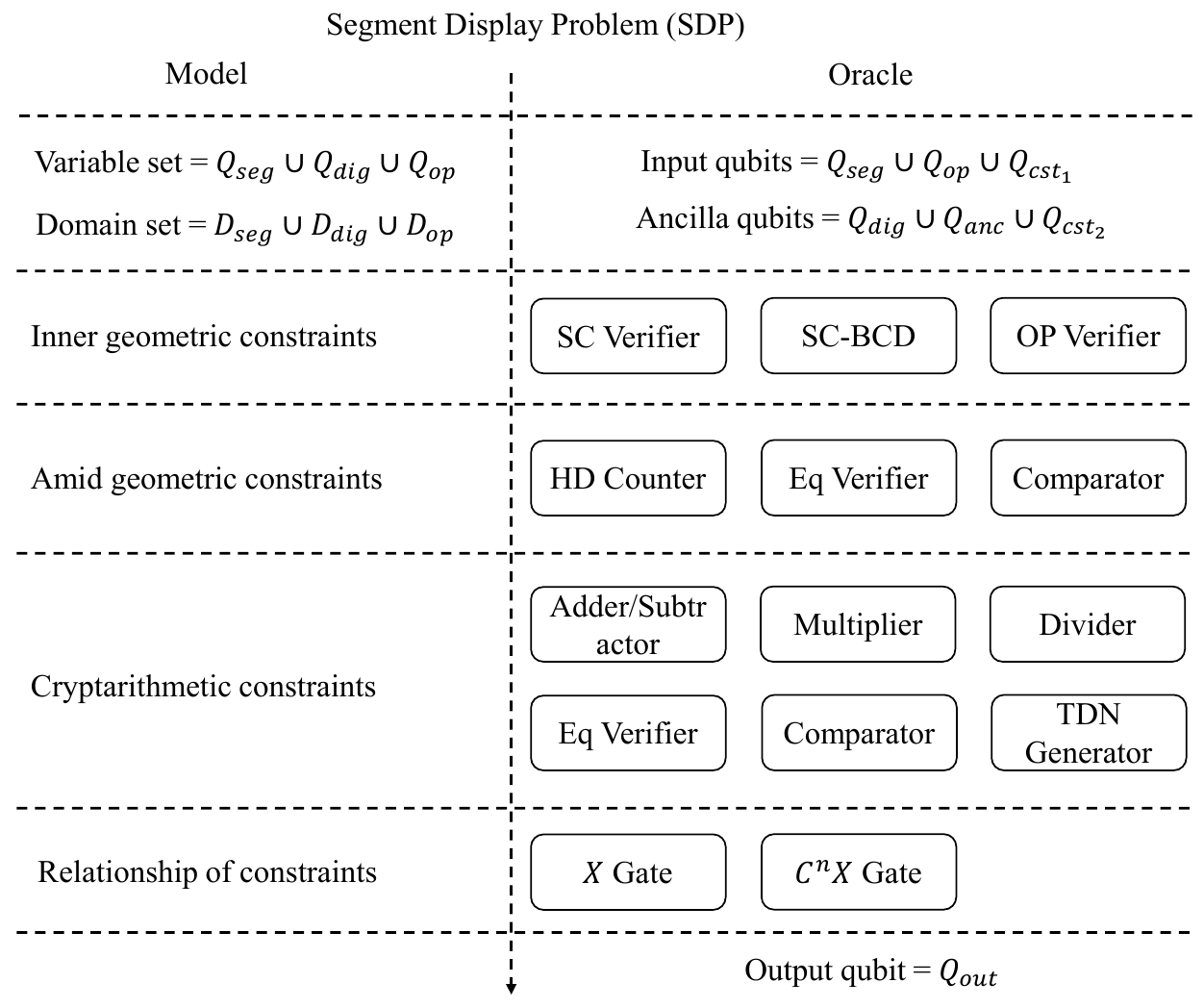}
    \caption{The relationship between the model and its Oracle for an SDP.}
    \label{fig:sdpflow}
\end{figure}

Given the composability of segment displays in an SDP, we propose designing the quantum oracle by combining several basic components, such as segment code (SC) verifiers, operation (OP) verifiers, and segment code-to-binary code decoders (SC-BCD). The relationship between the model and its quantum oracle for an SDP, as given in Figure~\ref{fig:sdpflow}, is explained in detail in the following content.

Based on the previous definitions of the variable sets $Q$ and $D$, the input qubits and ancilla qubits for a quantum oracle can be partially determined. Commonly, the input qubits consist of the elements in the variable set $Q_{seg}$ of physical segment displays and the set $Q_{op}$ of operations. Since the digit position variable $q_d$ is an abstract concept, the ancilla qubits can be determined by the variable set $Q_{dig}$ for digits and set $Q_{anc}$ for ancilla qubits to construct possible required intermediate variables when solving an SDP. Considering the limitation of the number of all qubits based on current quantum technology, the $Q_{seg}$ of the input qubits can be replaced with $Q_{dig}$, and the $Q_{dig}$ of ancilla qubits can be deleted. 

Moreover, according to whether the number $K$ (in Condition 4) is fixed in an SDP, the variable $K$ will be either the input qubits or the ancilla qubits of the oracle used to solve the SDP. The variable to denote the number $K$ is an element in the set $Q_{cst}$ of required variables to present constraints $C$. For this reason, the input qubits for a quantum oracle of an SDP can be the union of the set $Q_{seg}$, $Q_{op}$, and $Q_{cst_1}$, while the ancilla qubits be the union of $Q_{dig}$, $Q_{cst_2}$, and $Q_{anc}$ in Figure~\ref{fig:sdpflow}. The union of $Q_{cst_1}$ and $Q_{cst_2}$ is the set $Q_{cst}$. Notice that the difference between variables in the sets $Q_{cst_2}$ and $Q_{anc}$ is divided by the usage of the ancilla qubits, i.e., whether the ancilla qubits are required to describe the intermediate variables directly obtained from the constraint set $C$ or obtained from implementation of the quantum gates to obtain these intermediate variables. 

Then, we need to determine the required basic components of the oracle by the inner geometric constraints $C_{gin}$, amid geometric constraints $C_{gamid}$, and cryptarithmetic constraints $C_{art}$.

For the $C_{gin}$, the required basic components are determined by Theorem~\ref{th:gin_f}, which is related to the first condition and the meaning of an SC-State. Specifically, under the first condition, a component called the ``segment code (SC) verifier" is required to verify the correctness of SC-State, presenting the sparse fun $f_{gin}$ in the first relation of Theorem~\ref{th:gin_f}. Since an $n$-bit binary code is needed to represent the SC-State of each one-digit number or a letter, a segment code to a binary code decoder ``SC-BCD" is required to convert a $n$-bit binary code to a 4-bit binary code commonly used in classical computers, presenting the bijection fun $f_{gin}$ in the remained relation (despite the first) of Theorem~\ref{th:gin_f}. As the second condition is mentioned as a special case of the general $C_{gin}$, a component called the ``operator (OP) verifier" is needed to check the correctness of the operator, as shown in Figure~\ref{fig:sdpflow}.

For the $C_{gamid}$, the required basic components are determined by Theorem~\ref{th:gamid_f} related to the fourth condition. According to Theorem~\ref{th:gamid_f} and the number $K$ of supported changed segments in Condition 4, the primary concern is to check the number of changed segments by comparing the initial state of segment displays and operators in an SDP (like the example in Figure~\ref{fig:SDPexp}) and the generated final state of possible solutions. When the state of a segment changes, the values of two binary variables in different segment codes, denoted by $q_s$, change.
Therefore, the final Hamming distance (HD) of all SC-States and arithmetic operations becomes $2K$. 

Therefore, the required basic components for $C_{gamid}$ mainly are the following three components: (i) the ``Hamming distance (HD) counter" to calculate the HD between the SC-States of two $n$-segment displays, (ii) the ``equation (Eq) verifier" to check the equivalence between $2K$ and the HD obtained from (i), and (iii) the ``comparator" to compare the relation between the number $K$ and the HD obtained from (i), e.g., to determine which number is the lesser or greater number.

For the $C_{art}$, the required basic components are the arithmetic units in the quantum domain. The basic components include the adder, subtractor, multiplexer, divider, Eq verifier, comparator, etc. Most components for the $C_{art}$ are with respect to the operators in the domain $D_{op}$. Notably, the Eq verifier for $C_{art}$ is a similar component to that mentioned for $C_{gamid}$ in Figure~\ref{fig:sdpflow}, due to the equivalence between $2K$ and HD and the standard equivalence between two numbers.

Considering the third condition (also a $C_{art}$), another component is required to combine two $n$-segment displays. Take the example in Figure~\ref{fig:SDPexp}, there are two numbers on the right side of the SDP. Multiplication is implicitly performed on this side to generate a two-digit number. The component to combine two $n$-segment displays, which we termed as ``two-digit number (TDN) generator", is also used to extend the length of the digit number based on several $n$-segment displays by sequential combination along with some $C^2X$ gates (to realize the and relation).

So far, all essential basic components for constraints $C_{gin}$, $C_{gamid}$, and $C_{art}$ are introduced. We now need to consider the relationships among these constraints. Considering the SDP from an actual application, there is not only the AND relationship among these constraints, which can be implemented by $C^n{X}$ quantum gates ($n$ control qubits and one target qubit). For an SDP, a constraint may need to be unsatisfied while another constraint needs to be satisfied. Thus, the Pauli-X (X) gate is required to negate a constraint. Sometimes, a choice must be made among different constraints, i.e., the OR relations between constraints exist. The OR relation can be implemented by the combination of X gates and $C^nX$ gates. After the composition of X gates, $C^nX$ gates, and all previously mentioned basic components, the final output in $Q_{out}$ can be obtained. Notice that the final output is a single output qubit that was obtained from the whole oracle in Figure~\ref{fig:sdpflow}. Essentially, the oracle implements a composed function of all smaller functions of constraints that are implemented by the basic components.

\section{Methods for a Layout-Aware Logic Design}

In this section, we introduce the next level of our methodology, which is a layout-aware logic design of the three important above-mentioned components, i.e., the SC verifier, SC-BCD, and TDN generator. Other basic components not introduced in this section are the blocks that are usually other researchers' focus areas, such as the adder and subtractor. Notice that the comparator and Eq verifier in the design of the oracle are easily implemented by our proposed Stesso approach in~\cite{chen2025stesso}. Our Stesso approach is efficient for designing quantum circuits composed of $C^nX$ gates. All the used $C^nX$ gates in the design of the quantum circuit of all basic components are implemented by Stesso. For this reason, we will not introduce these two components or the details of generating a $C^nX$ gate in this paper.
In this section, we will specifically present the oracle design ideas of the SC verifier, SC-BCD, and TDN generator, along with their corresponding quantum circuits.

\subsection{Segment code verifier}

As stated in Theorem~\ref{th:gin_f}, an $n$-bit segment code (SC) verifier is exactly implemented by a sparse function $f_{gin}$ where most values are zero. Therefore, an SC verifier can be defined as the Definition~\ref{def:SSCV}.

\begin{definition}
    \label{def:SSCV}
    The segment code (SC) verifier is our proposed verifier that determines whether the SC-State displayed by the $n$-segment display represents a correct number between 0 and 9 or a correct letter.
    The input of this verifier is a segment variable $q_s = q_{s_1} \cdots q_{s_n}$, and the output is a single-bit binary variable $v_1$. The function that maps $q_s$ to $v_1$ is a sparse function.
\end{definition}

The basic circuit design idea of the SC verifier can be divided into three steps: (i) the first step is to present the SC-State circuits corresponding to the valid number or valid letters, respectively, based on the circuit composed of two $C^5X$ Toffoli gates implemented by Stesso structure in~\cite{chen2025stesso}, (ii) the second step is to find a combination sequence of correct SC-States corresponding to valid numbers or letters, such as $\{ s_i~|~0 \leq i \leq 9, i \in \mathbb{N} \}$, based on the Hamming distances among these correct SC-States. So that the sum of Hamming distances among all adjacent n-bit binary segment variables $q_s$ in this combination sequence is minimized, and (iii) the third step is to concatenate the circuits corresponding to the SC-States in a sequence based on this combination sequence, and then simplify the circuit.

Obviously, the key to designing the circuit for the SC verifier is to choose the appropriate combination sequence, which we termed the ``Combination Sequence of Exclusive Sums (CSES)".

Take the seven-segment display in Figure~\ref{fig:SDPexp} as an example, all correct SC-States are composed of ten SC-States, each representing a number from 0 to 9, as stated in Table~\ref{tab:ssd-state}. 
\begin{table}[h!]
    \centering
    \caption{All correct SC-States.}
    \begin{tabular}{c|c|c}
    \hline
       Symbol & Meaning & Segment code \textit{abcdefg}   \\
    \hline
       $s_0$ & Number 0 & $abcdef\bar{g}$ \\
       $s_1$ & Number 1 & $\bar{a}bc\bar{d}\bar{e}\bar{f}\bar{g}$ \\
       $s_2$ & Number 2 & $ab\bar{c}de\bar{f}g$ \\
       $s_3$ & Number 3 & $abcd\bar{e}\bar{f}g$ \\
       $s_4$ & Number 4 & $\bar{a}bc\bar{d}\bar{e}fg$ \\
       $s_5$ & Number 5 & $a\bar{b}cd\bar{e}fg$ \\
       $s_6$ & Number 6 & $a\bar{b}cdefg$ \\
       $s_7$ & Number 7 & $abc\bar{d}\bar{e}\bar{f}\bar{g}$ \\
       $s_8$ & Number 8 & \textit{abcdefg} \\
       $s_9$ & Number 9 & $abcd\bar{e}fg$ \\
       \hline
    \end{tabular}
    \label{tab:ssd-state}
\end{table}

Except for the ten correct SC-States in Table~\ref{tab:ssd-state}, all other $2^7 - 10$ possible SC-States are incorrect. This fact also partially proved that the description about the sparse function $f_{gin}$ is correct in Definition~\ref{def:SSCV} and Theorem~\ref{th:gin_f}.

Our methodology for choosing the appropriate CSES can be further divided into three steps as follows. Notice that, in this paper, a minterm means a product term of all variables, such as \textit{abcdefg}, which is exactly the segment variable $q_s$.

When the input variables \textit{abcdefg} of the SC Verifier are equal to any of the situations stated in Table 1, the value of the output variable $v_1$ is 1; for all other situations, the value of the output variable $v_1$ is 0, as expressed in Equation~(\ref{eq:SSCV}). The notation $\oplus$ represents a Boolean XOR operation.
\begin{equation}
    \label{eq:SSCV}
    \begin{aligned}
    v_1 =& s_0 \oplus s_1 \oplus s_2 \oplus s_3 \oplus s_4 \oplus s_5
           \oplus s_6 \oplus s_7 \oplus s_8 \oplus s_9
        \\=& abcdef\bar{g} \oplus
            \bar{a}bc\bar{d}\bar{e}\bar{f}\bar{g} \oplus
            ab\bar{c}de\bar{f}g \oplus
            abcd\bar{e}\bar{f}g \oplus
            \bar{a}bc\bar{d}\bar{e}fg \oplus \\&
            a\bar{b}cd\bar{e}fg \oplus 
            a\bar{b}cdefg \oplus 
            abc\bar{d}\bar{e}\bar{f}\bar{g} \oplus 
            abcdefg \oplus 
            abcd\bar{e}fg
    \end{aligned}
\end{equation}



\begin{table}[h!]
    \centering
    \caption{Number of all minterms of each variable in all correct SC-States. Notice that the occurrences mean the number of each variable in minterms.}
    \begin{tabular}{c|c|c}
    \hline
       Variable & Number of positive occurrences & Number of negative occurrences   \\
    \hline
       $a$ & 8 & 2 \\
       $b$ & 8 & 2 \\
       $c$ & 9 & 1 \\
       $d$ & 7 & 3 \\
       $e$ & 4 & 6 \\
       $f$ & 6 & 4 \\
       $g$ & 7 & 3 \\
       \hline
    \end{tabular}
    \label{tab:numterm}
\end{table}

\noindent\textbf{Step 1.}
Decide the initial minterm of the CSES.

Firstly, count the number of positive and negative occurrences of each variable from Table~\ref{tab:ssd-state}.
For example, the number of positive occurrences of variable $c$ in Table~\ref{tab:ssd-state} is 9, and the number of negative occurrences is 1.
Here, we list the number of all occurrences of each variable stated in Table~\ref{tab:numterm}.

In Table~\ref{tab:numterm}, we now list the positive and negative occurrences of each variable in decreasing order.
Therefore, the first item of the combination sequence is then obtained, i.e., $abcd\bar{e}fg$, that is exactly represents the correct SC-State $s_9$ of Number 9.
Because of the reconstruction of a positive or negative form of each variable from all correct SC-States, the initial minterm of this combination sequence is not necessarily one of the required correct SC-States.
For instance, when the decreasing order of occurrences of each variable is $\{ a, b, c, d, \bar{e}, f, \bar{g} \}$, the first minterm of the combination sequence becomes $abcd\bar{e}f\bar{g}$, which is not a correct SC-State representing a number.

\begin{figure}[htbp]
    \centering
    \includegraphics[width=0.85\linewidth]{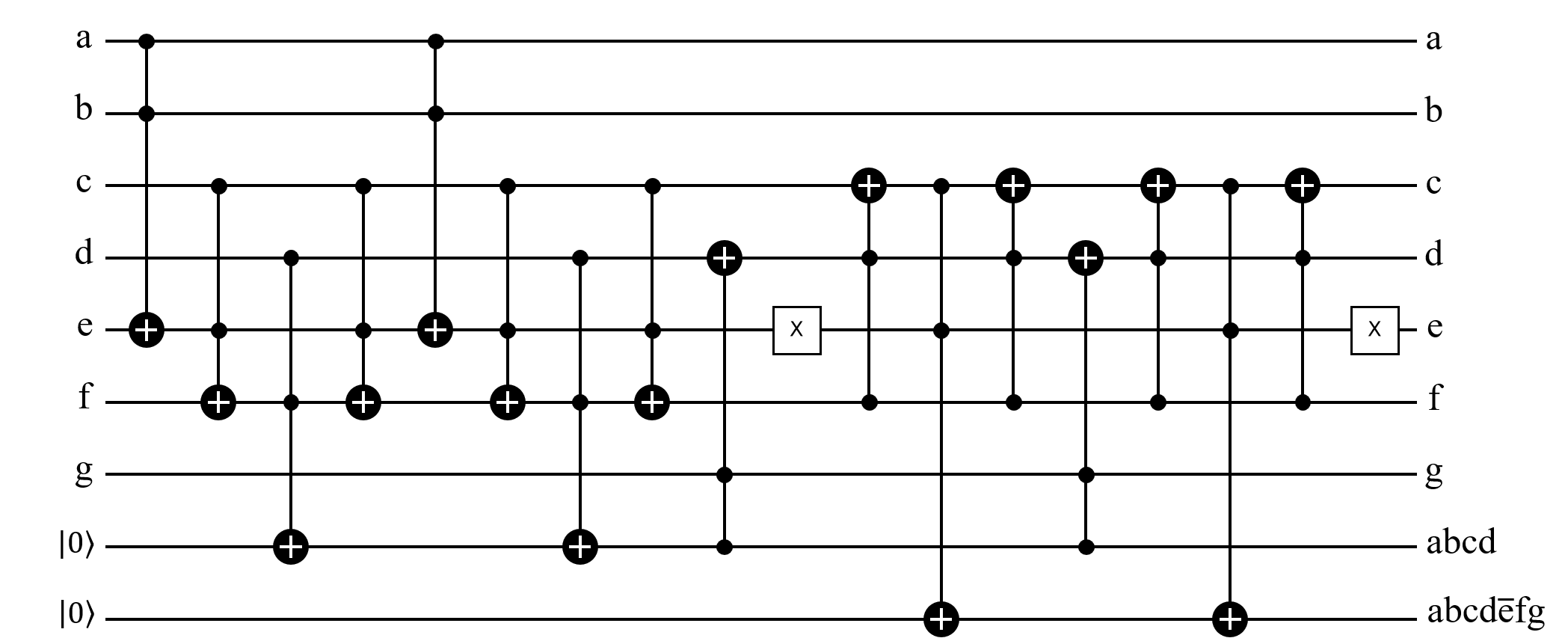}
    \caption{The SC-State of Number 9, which is an 8-bit Toffoli gate, where $a$, $b$, $c$, $d$, $e$, $f$, $g$ are controls, $c$, $d$, $e$, $f$, and the first ancilla qubit are the step-outputs, and the last ancilla qubit is the target.}
    \label{fig:SC-State-9}
\end{figure}

According to the initial minterm $abcd\bar{e}fg$, we now define the combination of two step-decreasing V-shaped logical structures to implement an 8-bit Toffoli gate.
Since the step-output of the first ancilla qubit is reusable, we choose the term $abcd$, which is mostly used in Table~\ref{tab:ssd-state}.
Therefore, the quantum circuit of SC-State of Number 9, which consists of two step-decreasing V-shaped logical structures and two inverters, is shown in Figure~\ref{fig:SC-State-9}.

\noindent\textbf{Step 2.}
Find all the following minterms of the CSES.

According to the initial minterm $abcd\bar{e}fg$, calculate the Hamming distance between the initial minterm $s_9$ and all other minterms, which indicate the remaining correct SC-State of Numbers 0 to 8, as stated in Table~\ref{tab:hd9}.
\begin{table}[h!]
    \centering
    \caption{The Hamming distance between $s_9$ and $s_i$ where $0 \leq i \leq 8, i \in \mathbb{N}$. Notice that any literal can be a positive or negative variable.}
    \begin{tabular}{c|c|c|c}
    \hline
       Minterm & HD & Different literals in $abcd$ & Different literals in $efg$   \\
    \hline
       $s_0$ & 2 &  & $e\bar{g}$ \\
       $s_1$ & 4 & $\bar{a} \bar{d} $ & $\bar{f}\bar{g}$ \\
       $s_2$ & 3 & $\bar{c}$ & $e\bar{f}$ \\
       $s_3$ & 1 & & $\bar{f}$ \\
       $s_4$ & 2 & $\bar{a} \bar{d} $ & \\
       $s_5$ & 1 & $\bar{b} $ & \\
       $s_6$ & 2 & $\bar{b} $ & $e$ \\
       $s_7$ & 3 & $\bar{d} $ & $\bar{f}\bar{g}$ \\
       $s_8$ & 1 & & $e$ \\
       \hline
    \end{tabular}
    \label{tab:hd9}
\end{table}

In this paper, we introduced two rules for choosing the next minterm $s_i$ for the combination sequence, where $0 \leq i \leq 8,~i\in \mathbb{N}$.
The first rule chooses the minterm $s_i$ with the minimum difference between $s_i$ and its previous minterm $s_{i-1}$ with variables $abcd$.
The second rule chooses the following minterm $s_i$ with the minimum difference between $s_i$ and its previous minterm $s_{i-1}$ with variables $efg$.
Notice that, always follow the first rule, then follow the second rule.
In this way, we should obtain the whole sequence that is an approximate minimum sum of Hamming distance between all neighboring minterms.
Based on Table~\ref{tab:hd9} using HD, the found combination sequence is $\{ s_9, s_8, s_0, s_3, s_2, s_6, s_5, s_7, s_1, s_4 \}$.\\


\noindent\textbf{Step 3.}
Merge some minterms of the CSES using a local transformation between two Stesso structures.

Based on the HD between each current minterm $s_i$ with its previous $s_{i-1}$ and next $s_{i+1}$ minterms, decide whether these minterms should be merged using XORs $(\oplus)$ gates or not.
If the merge between these minterms is required, we apply the local transformation of two step-decreasing V-shaped logical structures.
Here, the combination sequence is merged into a sequence $\{ s_9 \oplus s_8,  s_0, s_3 \oplus s_2, s_6 \oplus s_5, s_7 \oplus s_1 \oplus s_4 \}$.
The merging pattern is also decided by considering both HD and the reusable terms of the SC-BCD component.
All reusable terms will be illustrated in the description of the SC-BCD component.\\

After the above-presented three steps to obtain the final combination sequence, 
the corresponding output $v_1$ is simplified as stated in Equation (\ref{eq:SSCV_T}).
Notice that the final combination sequence is in the form of Combination Sequence of Exclusive Sums (CSES), which we introduced in this paper as a useful concept for the layout-aware logical structural synthesis utilized for real quantum computers.
\begin{equation}
    \label{eq:SSCV_T}
    \begin{aligned}
    v_1 =& (s_9 \oplus s_8) \oplus s_0 \oplus (s_3 \oplus s_2) \oplus (s_6 \oplus s_5) \oplus (s_7 \oplus s_1 \oplus s_4)
        \\=& 
        (abcd\bar{e}fg \oplus abcdefg) \oplus
        abcdef\bar{g} \oplus \\&
        (abcd\bar{e}\bar{f}g \oplus ab\bar{c}de\bar{f}g) \oplus
        (a\bar{b}cdefg \oplus a\bar{b}cd\bar{e}fg) \oplus \\&
        (abc\bar{d}\bar{e}\bar{f}\bar{g} \oplus
        \bar{a}bc\bar{d}\bar{e}\bar{f}\bar{g} \oplus
        \bar{a}bc\bar{d}\bar{e}fg)
        \\=& abcdfg \oplus
             abcdef\bar{g} \oplus
             abd\bar{f}g(c \oplus e) \oplus
             a\bar{b}cdfg \oplus
             (bc\bar{d}\bar{e}\bar{f}\bar{g} \oplus
             \bar{a}bc\bar{d}\bar{e}fg)
    \end{aligned}
\end{equation}



\begin{figure}[htbp]
    \centering
    \begin{subfigure}[b]{1\textwidth}
        \centering
        \includegraphics[width=1\textwidth]{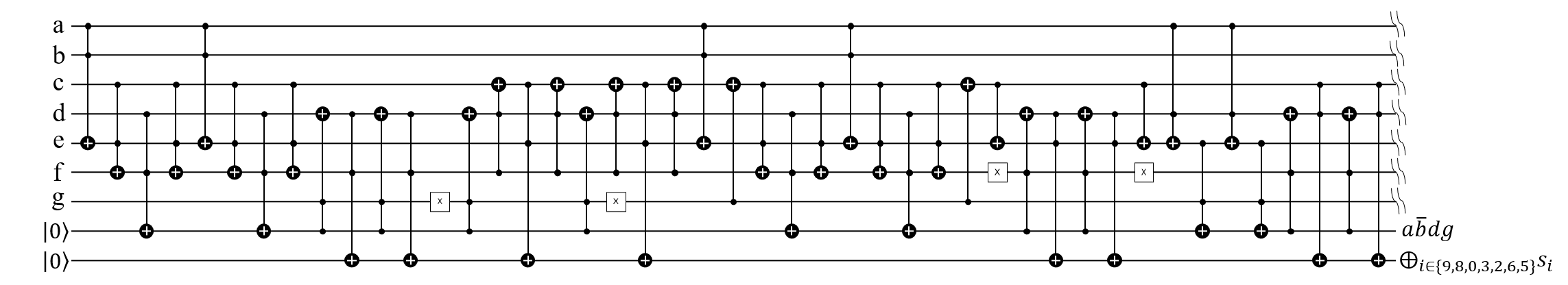}
        \caption{First part}
        \label{fig:SSC-Verf-P1}
    \end{subfigure}
    \vfill
    \vspace{0.05\textwidth}
    \begin{subfigure}[b]{1\textwidth}
        \centering
        \includegraphics[width=1\textwidth]{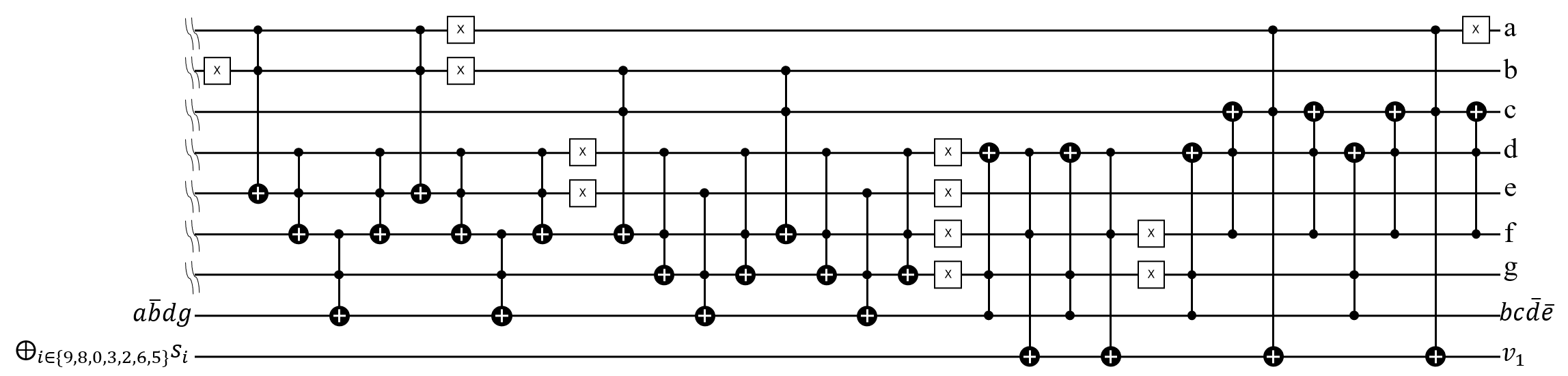}
        \caption{Second part}
        \label{fig:SSC-Verf-P2}
    \end{subfigure}
    \caption{Compositional parts of the SC Verifier of seven-bit binary segment variable $q_s = abcdefg$, i.e., seven input qubits $a,b,c,d,e,f,g$, one step-output qubit, and one output qubit $v_1$.}
    \label{fig:SSC-Verf}
\end{figure}

Notably, the quantum circuits of all correct SC-States can not be directly combined because of the reusable step-output of an SC-State. In Equation~(\ref{eq:SSCV_T}), the step-output of an SC-State required for the CSES has been updated in the order of $\{abcd, abdg, a\bar{b}dg, bc\bar{d}\bar{e}\}$. Then, we obtain the quantum circuit of the SC Verifier in Figure~\ref{fig:SSC-Verf}.

In this paper, the quantum cost is defined as the total number of required qubits and quantum gates in the quantum oracle. In Figure~\ref{fig:SSC-Verf}, the total number of qubits is nine (seven inputs, one step-output, and one output), and the total number of gates is 88 (68 standard 3-bit Toffoli $C^2X$ gates, four Feynman $CX$ gates, and 16 $X$ gates).

\subsection{Segment code to binary code decoder}

In general, an $n$-bit binary code with a smaller value of $n$ has a simpler circuit design for arbitrary operations, no matter whether it is used in classical or quantum computers. 
Specifically, if we directly use the SC $q_s$ that indicates a number in our SDP, the operational complexity of quantum circuit design is dramatically increased compared with a 4-bit binary code $q_d$.
Therefore, we design a segment code to binary code decoder (SC-BCD) in Definition~\ref{def:sscbcd}.

\begin{definition}
    \label{def:sscbcd}
    The segment code-to-binary code decoder (SC-BCD) is our proposed decoder that converts the segment code $q_s$ to the binary code $q_d$, which represents the digits 0 to 9. The input of this decoder is an $n$-bit binary code $q_s= q_{s_1} \cdots q_{s_n}$, e.g., $q_s = abcdefg$ for a seven-segment display, and its output is usually a 4-bit binary code variable $q_d = x_1 x_2 x_3 x_4$.
\end{definition}

Still, take a seven-segment display as an example, we list the corresponding relation between the segment code (SC) and binary code (BC) for each Number from 0 to 9, as stated in Table~\ref{tab:ssc2BCD}.

\begin{table}[htbp]
    \centering
    \caption{Relations between SC and bC.}
    \begin{tabular}{c|c|c}
    \hline
       Number & SC $abcdefg$  & BC $x_4 x_3 x_2 x_1$ \\
    \hline
       0 & 1111110  & 0000 \\
       1 & 0110000 & 0001 \\
       2 & 1101101 & 0010 \\
       3 & 1111001 & 0011 \\
       4 & 0110011 & 0100 \\
       5 & 1011011 & 0101 \\
       6 & 1011111 & 0110 \\
       7 & 1110000 & 0111 \\
       8 & 1111111 & 1000 \\
       9 & 1111011 & 1001 \\
       \hline
    \end{tabular}
    \label{tab:ssc2BCD}
\end{table}

Note that the output $x_4 x_3 x_2 x_1$ of the SC-BCD only works under the condition that the output value $v_1$ of the SC Verifier is equal to $1$.
Because $v_1 = 1$ means that there are only ten SC-States for all numbers remaining, all the rest $(2^7 - 10)$ possible SC-States with no correct meaning cannot occur in this case.

According to the relations from Table~\ref{tab:ssc2BCD} and all correct SC-States from Table~\ref{tab:ssd-state}, the value of a 4-bit binary code $q_{d_4}q_{d_3}q_{d_2}q_{d_1}$ is equal to the combination of some minterms $s_i$, as expressed in Equation (\ref{eq:BCD}).
\begin{equation}
    \label{eq:BCD}
    \begin{aligned}
        {x_4} &= s_8 \oplus s_9
        \\
        {x_3} &= s_4 \oplus s_5 \oplus s_6 \oplus s_7
        \\
        {x_2} &= s_2 \oplus s_3 \oplus s_6 \oplus s_7
        \\
        {x_1} &= s_1 \oplus s_3 \oplus s_5 \oplus s_7 \oplus s_9
    \end{aligned}
\end{equation}

Since the outputs $x_4, x_3, x_2, x_1$ of the decoder consist of the same minterms $s_i$, where $1 \leq i \leq 9$, we can reuse some partial circuits of the SC Verifier.
Comparing the CSES of $\{ s_9 \oplus s_8,  s_0, s_3 \oplus s_2, s_6 \oplus s_5, s_7 \oplus s_1 \oplus s_4 \}$ for the SC Verifier with Equation (\ref{eq:BCD}), some single minterms (even some combined terms), are reusable. For instance, the terms $s_9 \oplus s_8$, $s_3 \oplus s_2$ and $s_6 \oplus s_5$ are both reusable.

The value of the XOR operation between two same terms is equal to $0$.
While the value of the XOR operation between $0$ and another term is equal to the term itself.
Therefore, the method to reuse the terms of the CSES $\{ s_9 \oplus s_8,  s_0, s_3 \oplus s_2, s_6 \oplus s_5, s_7 \oplus s_1 \oplus s_4 \}$ is to add Feynman gates at different positions in the circuit of the SC Verifier.
For example, $0 \oplus (s_9 \oplus s_8) = s_8 \oplus s_9 = x_4$ and $0 \oplus ((s_9 \oplus s_8) \oplus s_0) \oplus ((s_9 \oplus s_8) \oplus s_0 \oplus (s_3 \oplus s_2)) = s_2 \oplus s_3$.

\begin{figure}[htbp]
    \centering
    \begin{subfigure}[b]{1\textwidth}
        \centering
        \includegraphics[width=1\textwidth]{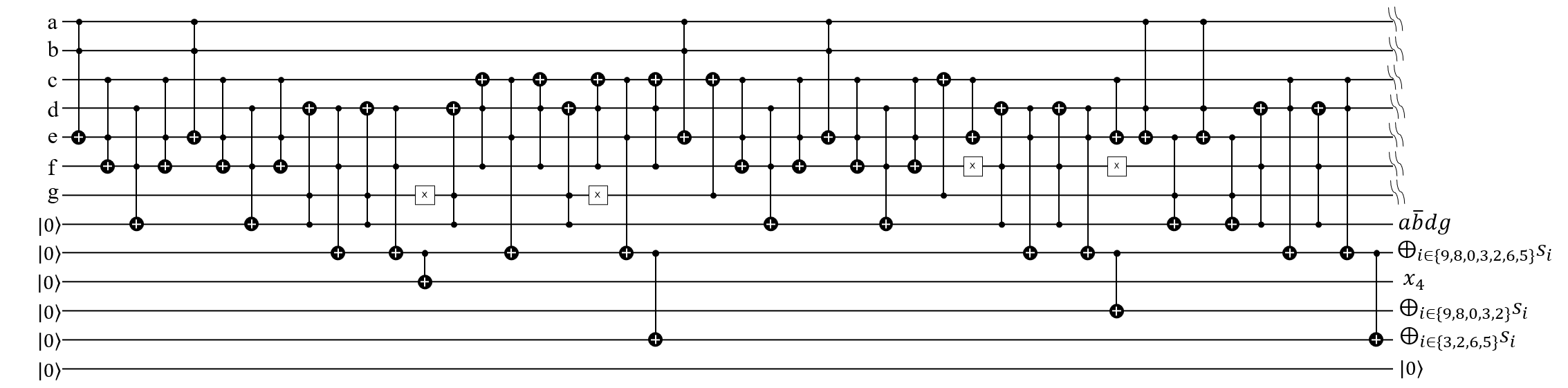}
        \caption{First part}
        \label{fig:SC-BCD-P1}
    \end{subfigure}
    \vfill
    \vspace{0.05\textwidth}
    \begin{subfigure}[b]{1\textwidth}
        \centering
        \includegraphics[width=1\textwidth]{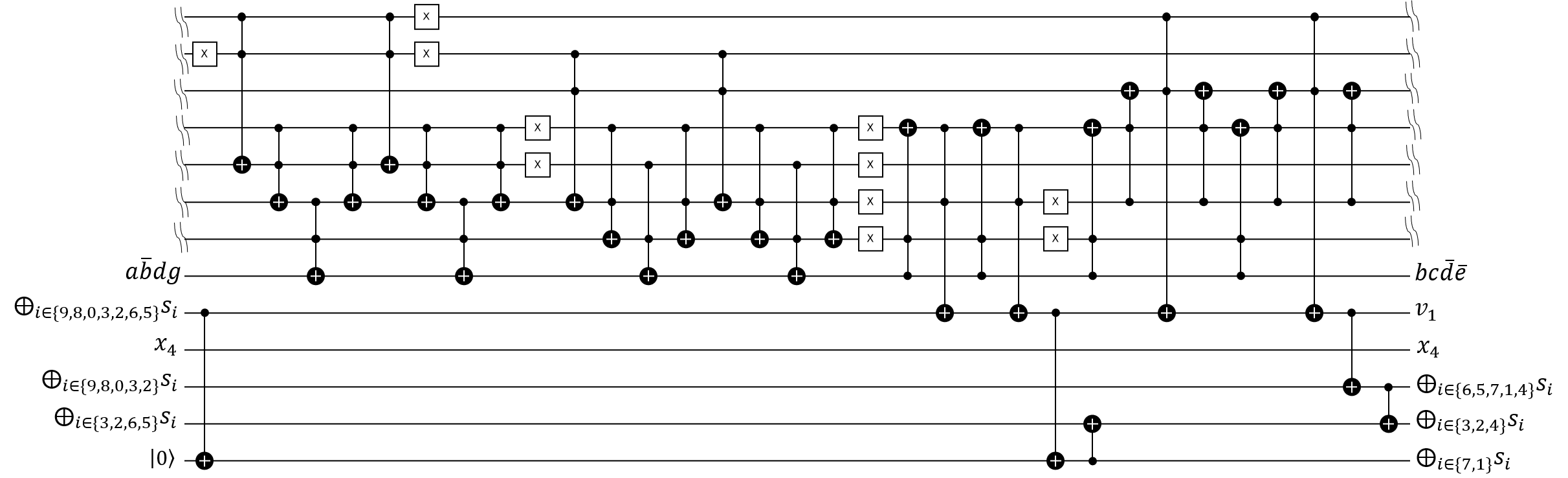}
        \caption{Second part}
        \label{fig:SC-BCD-P2}
    \end{subfigure}
    \vfill
    \vspace{0.05\textwidth}
    \begin{subfigure}[b]{1\textwidth}
        \centering
        \includegraphics[width=1\textwidth]{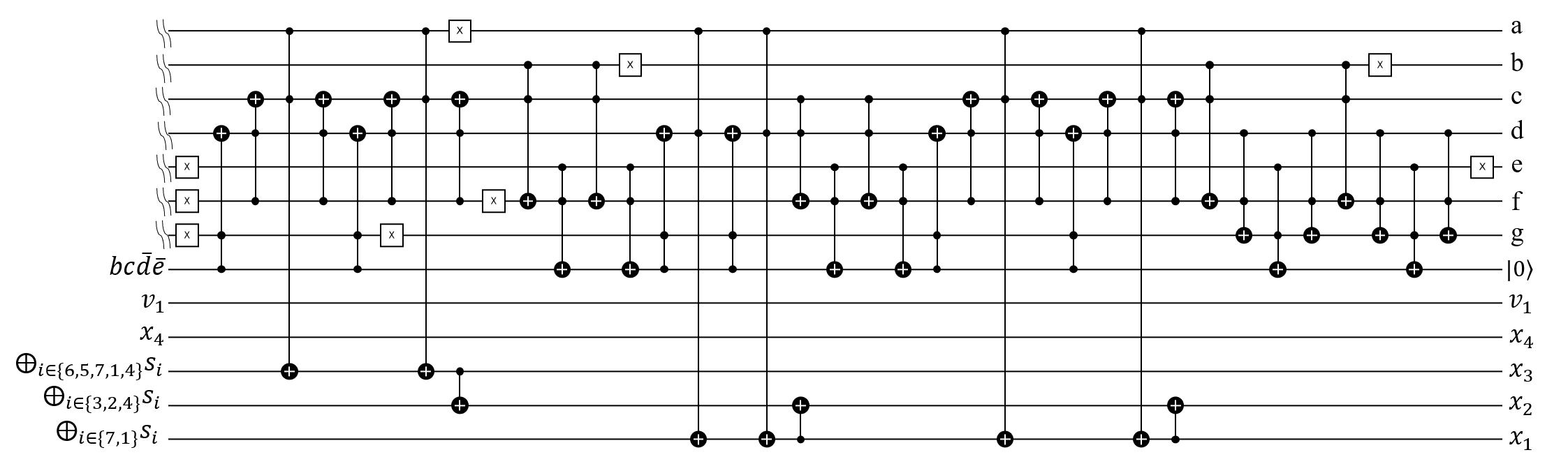}
        \caption{Third part}
        \label{fig:SC-BCD-P3}
    \end{subfigure}
    \caption{Compositional parts of the combination circuit of SC Verifier and SC-BCD of seven-bit binary segment variable $q_s = abcdefg$, i.e., seven input qubits $a,b,c,d,e,f,g$, one step-output qubit, and variable $v_1$ and digit position variable $q_d = x_1 x_2 x_3 x_4$, i.e., five output qubits $v_1$, $x_4$, $x_3$, $x_2$, $x_1$.}
    \label{fig:SC-BCD}
\end{figure}

Based on the method of reusing terms related to $s_i$, we only need to add some Feynman gates.
Compared to directly combining the circuits of all required $s_i$ from Equation (\ref{eq:BCD}), the cost of this method to implement the SC-BCD is much lower. However, this also makes it difficult to separate the SC-BCD and SC Verifier.
Using the method of reusing minterms $s_i$ and the output $v_1$ in Equation (\ref{eq:SSCV}), the outputs $x_4 x_3 x_2 x_1$ in Equation (\ref{eq:BCD}) can be transformed into Equation (\ref{eq:BCD_T}).
\begin{equation}
    \label{eq:BCD_T}
    \begin{aligned}
        {x_4} &= s_8 \oplus s_9
        \\
        {x_3} &= s_4 \oplus s_5 \oplus s_6 \oplus s_7
            = (s_9 \oplus s_8 \oplus s_0 \oplus s_3 \oplus s_2) \oplus v_1 \oplus s_1
        \\
        {x_2} &= s_2 \oplus s_3 \oplus s_6 \oplus s_7
            = ((s_3 \oplus s_2) \oplus (s_6 \oplus s_5)) \oplus (s_7 \oplus s_1) \oplus s_1 \oplus s_5
        \\
        {x_1} &= s_1 \oplus s_3 \oplus s_5 \oplus s_7 \oplus s_9
            = (s_7 \oplus s_1) \oplus (s_9 \oplus s_3) \oplus s_5
    \end{aligned}
\end{equation}
Therefore, the combined circuit of the SC Verifier and the SC-BCD is shown in Figure~\ref{fig:SC-BCD}. In Figure~\ref{fig:SC-BCD}, the cost of gates for this SC-BCD is 56 (36 standard 3-bit Toffoli $C^2X$ gates, 12 Feynman $CX$ gates, and 8 $X$ gates).


\subsection{Two-digit number generator}

Considering the right side of the cryptarithmetic equation in Figure~\ref{fig:SDPexp}, the two-digit number (TDN) generator is required to combine the two numbers, as described in Definition~\ref{def:twodig}.

\begin{definition}
    \label{def:twodig}
    The two-digit number (TDN) generator is our proposed generator to convert two one-digit numbers into a two-digit number. The inputs of this TDN generator are two 4-bit binary codes ($x_4 x_3 x_2 x_1$ and $y_4 y_3 y_2 y_1$), and its output is an 8-bit binary code $z_8 z_7 z_6 z_5 z_4 z_3 z_2 z_1$.
\end{definition}

In the two-digit number generator, the $x_4 x_3 x_2 x_1$ represents the tens place number, and the $y_4 y_3 y_2 y_1$ represents the units place number. Therefore, the two-digit number ($z_8 z_7 z_6 z_5 z_4 z_3 z_2 z_1$) is the sum of the number ($y_4 y_3 y_2 y_1$) and the multiplication result between the number ($x_4 x_3 x_2 x_1$) and ten.


Since the multiplication involved in generating a two-digit number is the multiplication between $x_4 x_3 x_2 x_1$ and $(1010)_2$, the computational process of deriving a two-digit number is simplified.

Then, the step carries $c_i$, step outputs $p_i$, final carries $c'_i$, and final outputs $z_i$ can be calculated in Equation (\ref{eq:2digit1}) and Equation (\ref{eq:2digit2}), respectively.
\begin{equation}
    \label{eq:2digit1}
    \begin{aligned}
        c_4 &= x_3 x_1, \;\; c_5 = x_4 x_2 \oplus c_4 (x_4 \oplus x_2), \;\;
        c_6 = c_5 x_3, \;\; c_7 = c_6 x_4\\
        p_4 &= x_3 \oplus x_1, \;\;
        p_5 = x_4 \oplus x_2 \oplus c_4, \;\;
        p_6 = x_3 \oplus c_5, \;\;
        p_7 = x_4 \oplus c_6, \;\;
        p_8 = c_7
    \end{aligned}
\end{equation}
\begin{equation}
    \label{eq:2digit2}
    \begin{aligned}
        c'_2 &= x_1 y_2, \;\;
        c'_3 = x_2 y_3 \oplus c'_2(x_2 \oplus y_2), \;\;
        c'_4 = p_4 y_4 \oplus c'_3(p_4 \oplus y_4),\\
        c'_5 &= c'_4 p_5, \;\;
        c'_6 = c'_5 p_6, \;\;
        c'_7 = c'_6 p_7, \;\;
        c'_8 = c'_7 p_8 \\
        z_1 &= y_1, \;\; z_2 = x_1 \oplus y_2, \;\;
        z_3 = x_2 \oplus y_3 \oplus c'_2, \;\;
        z_4 = p_4 \oplus y_4 \oplus c'_3, \\
        z_5 &= p_5 \oplus c'_4, \;\;
        z_6 = p_6 \oplus c'_5, \;\;
        z_7 = p_7 \oplus c'_6, \;\;
        z_8 = p_8 \oplus c'_7, \;\;
        z_9 = c'_8
    \end{aligned}
\end{equation}

Since the $x_4 x_3 x_2 x_1$ represents a one-digit number from zero to nine, the step output $p_8 = c_7 = c_6 x_4 = 0$, and the $p_8, p_7, p_6, p_5, p_4$ cannot be equal to 1 simultaneously.
For this reason, the output $z_9 = c'_8 = 0$ and $z_8 = c'_7$.
So the output of the two-digit number generator can be coded as an 8-bit binary code, not a 9-bit binary code, which corresponds to the representative range of a two-digit number from 00 to 99.

Based on the above facts, combining two binary code adder circuits can construct a two-digit number generator circuit. Since the result $p_7 p_6 p_5 p_4$ of the previous adder circuit is the input of the subsequent adder circuit, the circuit obtained by directly combining two adder circuits is not the most efficient circuit design for a two-digit number generator, due to the inputs of two adders can not be overlapped on the same wire as well as adding more ancilla qubits increases the final quantum cost.

According to Equation (\ref{eq:2digit2}) and the range of a two-digit number (00 to 99), the intermediate result of the two-digit number generator circuit includes a total of three step-carries $c_4, c_5, c_6$, four step-outputs $p_4, p_5, p_6, p_7$, and six final carries $c_2', c_3', c_4', c_5', c_6', c_7'$. Among them, three step-outputs $p_5, p_6, p_7$ are implemented based on three step-carries $c_4, c_5, c_6$. Four step-outputs $p_4, p_5, p_6, p_7$ are the inputs of the second adder. Six carries $c_2', c_3', c_4', c_5', c_6', c_7'$ are the intermediate outputs of the second adder and also the intermediate inputs of the $z_i$ output of the second adder. 

\begin{figure}[htbp]
    \centering
    \includegraphics[width=0.95\linewidth]{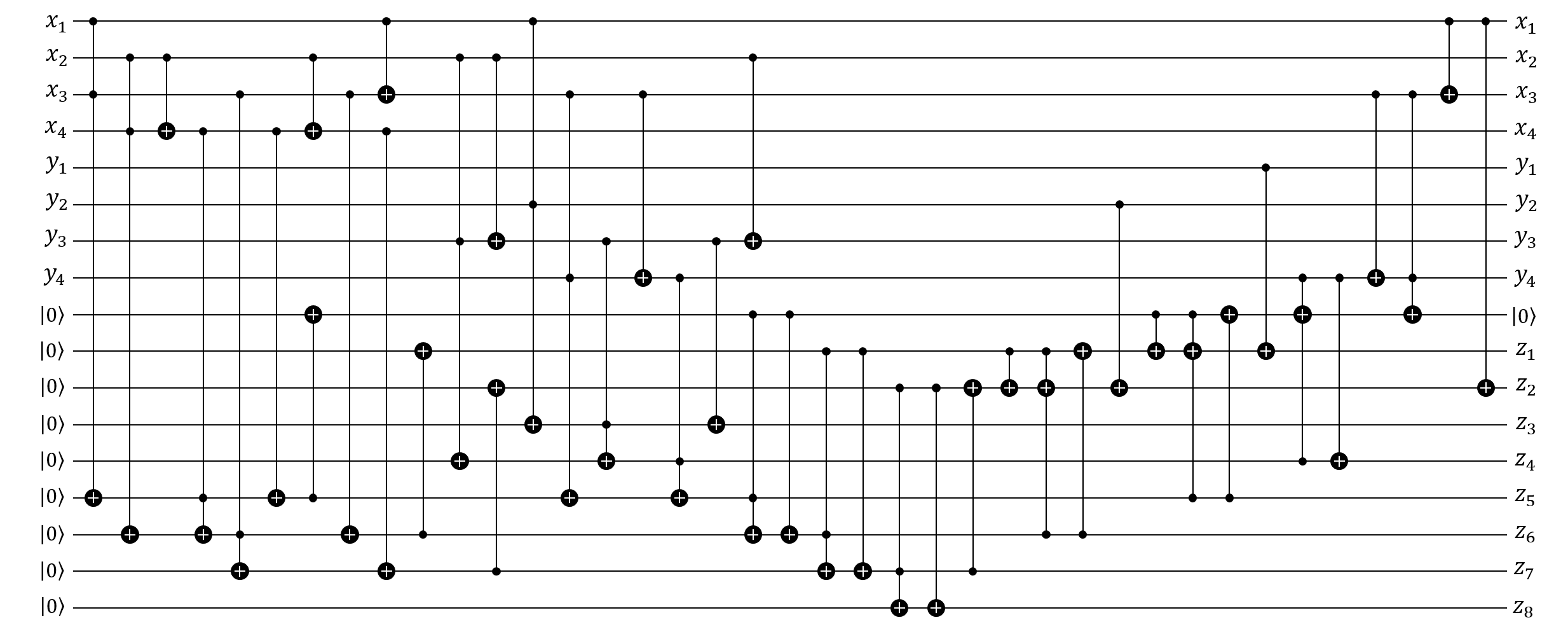}
    \caption{The quantum circuit of the two-digit number generator.}
    \label{fig:2digit-gen}
\end{figure}

In this paper, our design methodology aims to reduce the number of required ancilla qubits and quantum gates by implementing the final outputs in the order of $\{ z_5, z_6, z_7, z_8, z_4, z_3, z_2, z_1\}$, as we termed them ``stages". Specifically, we utilized the first three ancilla qubits with an initial state of $\ket{0}$ to represent $p_5, p_6, p_7$. After implementing the circuit from $z_5$ to $z_8$, we re-represented the stages of the first three ancilla qubits into the $\ket{0}$ state to facilitate the subsequent circuit design from $z_1$ to $z_4$. The circuit of a two-digit number generator can be implemented by the binary code adder, as illustrated in Figure~\ref{fig:2digit-gen}.

In Figure~\ref{fig:2digit-gen}, the total number of required qubits is 17 (eight inputs, one ancilla, eight outputs), and the total number of gates is 43 (16 standard 3-bit Toffoli $C^2X$ gates and 27 Feynman $CX$ gates).

\section{Case: a matchstick problem}

Based on the above-mentioned oracle design methodology, we applied it to obtain the whole quantum circuit of the oracle for solving a matchstick problem.
Due to the hardware limitation of the number of supported qubits in the current quantum simulator, we designed a quantum reversible circuit for amid geometric constraints $C_{gamid}$ (consisting of HD counter, Eq verifier, etc) to reuse 29 ancilla qubits and 30 input qubits. 

In general, the number of changed segments in a matchstick problem is typically one or two.
However, in this case, the number of supported changed segments $K$ in the fourth condition is smaller than eight, $K < 8$.
Notice that the constraint limitation of $K < 8$ is chosen due to the maximum number of changeable segments for a Number in a seven-segment display to be seven, the maximum number of changeable segments for an operation to be one, and the increasing evaluation complexity.

Here, Table~\ref{tab:quantum-cost} lists the quantum cost of all basic components for the entire oracle, including the total number of required qubits (input qubits, ancilla qubits, and output qubits) and quantum gates (3-bit Toffoli gates, Feynman gates, and NOT ``X'' gates).

\begin{table}[h!]
    \centering
    \caption{The quantum cost of all components for the entire oracle. Notations ``in'' as input qubits, ``anc'' as ancilla qubits, `` out'' as output qubits, ``$C^2X$'' as a 3-bit Toffoli gate, ``$CX$'' as a Feynman gate, and ``$X$'' as a Pauli-X gate.}
    \begin{tabular}{p{2cm}|cccc|cccc}
    \hline
    {Name} & \multicolumn{4}{c|}{Number of qubits} & \multicolumn{4}{c}{Number of quantum gates} \\ \hline
    & in & anc & out & total & $C^2X$ & $CX$ & $X$ & total \\ \hline
    SC verifier & 
    7 & 1 & 1 & 9
    & 68 & 4 & 16 & 88 \\ \hline
    SC verifier + SC-BCD & 
    7 & 1 & 5 & 13
    & 104 & 16 & 24 & 144 \\ \hline
    Adder / Subtractor & 
    10 & 1 & 5 & 16
    & 9 & 24 & 2 & 35 \\ \hline
    TDN generator & 
    8 & 1 & 8 & 17
    & 16 & 27 & 0 & 43 \\ \hline
    Eq verifier & 
    13 & 1 & 1 & 15
    & 16 & 10 & 16 & 42 \\ \hline
    SC-HDC & 
    14 & 2 & 3 & 19
    & 14 & 22 & 2 & 38 \\ \hline
    $n$-bit adder & 
    $2n$ & 0 & $n + 1$ & $3n + 1$
    & $2n - 1$ & $3n - 1$ & 0 & $5n - 2$ \\ \hline
    Oracle & 
    33 & 30 & 1 & 64 
    & 640 & 216 & 154 & 1010 \\
    \hline
    \end{tabular}
    \label{tab:quantum-cost}
\end{table}

To utilize Grover's quantum search algorithm, we combined our quantum circuit of the oracle and the controlled-diffuser in \cite{al2024concept}, to generate the entire circuit for simulation.
Specifically, we simulated our circuit using a simulated noisy model ``Qiskit-Aer" based on IBM ``torino" QPU, with a total number of supporting qubits is 133.

{For solving an exact matchstick puzzle, the initial state of the matchstick equation is usually fixed.
So, the number of all possible positions for a matchstick is 30, which equals the number of required input qubits that represent the initial state of four 7-bit seven-segment codes and one 2-bit operator code.
When the number of all matchsticks is $M_s$ and the number of movable matchsticks is $K$, the size of all possible configurations (as the search space) is $C_{30 - M_s + K}^{K} = \frac{(30 - M_s + K)!}{K!(30 - M_s)!}$, where $0 \leq K \leq M_s < 30$ and $K < 8$.}


Thus, we used one initial state of the matchstick equation as a simulation example in Figure~\ref{fig:SDPexp}.
The corresponding solutions of this equation are $3 + 6 = 09$ and $6 + 3 = 09$, i.e., the final state of solution consisting of $s_3, s_6, s_0, s_9$. The exact number of minimum-changed segments $K$ is 2.






\section{Conclusions and Future Work}

The Segment Display Problem (SDP), as a well-known computational problem and constraint satisfaction problem (CSP), includes matchstick and alphanumeric problems, in which such problems are not so simple to find solutions \cite{grabarchuk2008simple}.
These problems not only explain mathematical concepts \cite{danesi2002puzzle}, but also the approaches for finding their solutions are beneficial methods to human intelligence or the development of artificial intelligence, which certainly have enlightening significance in teaching and academia \cite{newell1959report}.

In the classical domain, there are many techniques to solve such SDPs, such as human deduction, heuristic search, and methods for solving Boolean satisfiability (SAT) and CSPs that are based on different problem design models.
The SDPs use visual representations of digits based on the $n$-segment display, where $n$ denotes the number of segments, typically seven, fourteen, or sixteen. The complexity of solving SDPs, which is represented by $m$ $n$-segment displays, significantly rises when $m$ and $n$ increase, where $m \geq 1$ and $n \geq 7$.

Even certain types of SDPs are known to be NP-hard or PSPACE-complete~\cite{uehara2023computational}; therefore, finding an optimal solution can be computationally very difficult, especially when the problem size increases. 

The goal of this paper is to introduce a new methodology for constructing SDPs and solving them using Grover's search algorithm in the quantum domain.
This newly introduced methodology for the oracle design is divided into a high-level and a logic level, where the high-level uses several of our designed basic components, and the logic level introduces the overall layout-aware design for the high-level components.
Our main contributions to the oracle design are:
(i)
defining the involving geometric and cryptarithmetic constraints of an SDP and mapping these constraints into different basic quantum components,
and (ii) inventing a new layout-aware logical structural synthesis, termed the ``Combination Sequence of Exclusive Sums (CSES)", to decrease the number of required quantum gates in the oracle.

Due to the composability of $n$-segment displays and the similarity between segment displays and other display technologies, our methodology is extendable for other CSPs. Specifically, the general design of CSES is suitable for other functions with only a few minterms, and all the basic components we introduce (which correspond to geometric and cryptarithmetic constraints in an SDP) are reusable and can be easily extended.
To demonstrate that our methodological approach can be used to implement and solve various SDPs successfully, we
utilize Grover’s quantum search algorithm with a noisy simulated model of IBM quantum computers, to experimentally search for solutions successfully on an example matchstick problem (as an SDP).


Future work toward our methodology will be to develop a quantum library comprising all components proposed in this paper. Therefore, the components of this library are reusable for other problem types. Another future work will extend and automate the category of applicable solvable problems in the quantum domain, such as alphabet-based problems, using constraint programming languages \cite{rossi2006handbook}, including Prolog \cite{clocksin2003programming,wielemaker2012swi,al2025prog} and Constraint Logic Programming (Real) ``CLP(R)" \cite{jaffar1987constraint}.






\bibliographystyle{ieeetr}
\bibliography{ref}

\end{document}